\documentclass[prd,twocolumn,showpacs]{revtex4}

\def\beq{\begin{equation}} \def\eeq{\end{equation}}
\def\bea{\begin{eqnarray}} \def\eea{\end{eqnarray}}

\preprint{\vbox{\baselineskip=12pt \rightline{gr-qc/0212126}
\rightline{ICN-UNAM-02/09}}}

\begin{document}

\title{On Quasinormal Modes,\\
 Black Hole Entropy, and Quantum Geometry}

\author{Alejandro Corichi}\email{corichi@nuclecu.unam.mx}
\affiliation{Instituto de Ciencias Nucleares\\
Universidad Nacional Aut\'onoma de M\'exico\\
A. Postal 70-543, M\'exico D.F. 04510, MEXICO}


\begin{abstract}
Loop quantum gravity can account for the Bekenstein-Hawking
entropy of a black hole provided a free parameter is chosen
appropriately. Recently, it was proposed that a new choice of the
Immirzi  parameter could predict both black hole entropy and the
frequencies of quasinormal modes in the large $n$ limit, but at
the price of changing the gauge group of the theory. In this note
we use a simple physical argument within loop quantum gravity to
arrive at the same value of the parameter.  The argument uses
strongly the necessity of having fermions satisfying basic
symmetry and conservation principles, and therefore supports SU(2)
as the relevant gauge group of the theory.
\end{abstract}
\maketitle

\section{Introduction}

Loop quantum gravity (LQG) has become in the past years a serious
candidate for a non-perturbative quantum theory of gravity
\cite{LQG}. Its most notorious predictions are the quantization of
geometry \cite{quantgeom} and the computation of black hole
entropy \cite{entropy,abck}. One of its shortcomings is the
existence of a one parameter family of inequivalent quantum
theories labelled by the Immirzi parameter $\gamma$
\cite{immirzi}. The black hole entropy calculation was proposed as
a way of fixing the Immirzi parameter $\gamma$ (and thus the
spectrum of the geometric operators) \cite{ac:kk}. When a
systematic approach to quantum black hole entropy was available
\cite{abck}, this was used to fix the value of $\gamma$ to be,
 \beq
      \gamma=\frac{\ln(2j_{\rm min}+1)}{2\pi\sqrt{j_{\rm
      min}(j_{\rm min}+1)}}
 \eeq
where $j_{\rm min}$ is the minimum (semi-integer) label for the
representations of SU(2), responsible for the entropy of the black
hole. At that time the most natural assumption was that $j_{\rm
min}=1/2$, giving an Immirzi parameter $\gamma_{\rm
abck}=\frac{\ln(2)}{\pi\sqrt{3}}$ \cite{abck}. Recently, Dreyer
made the bold suggestion that there is an independent way of
fixing the Immirzi parameter \cite{dreyer}, using very little
information about LQG. The new approach is based on a conjecture
by Hod that the quasinormal mode frequencies $\omega_{\rm QNM}$,
for large $n$ have an asymptotic behavior given by \cite{hod},
 \beq
  M\omega_{\rm QNM}=\frac{\ln 3}{8\pi}\label{Hod}
 \eeq
This conjecture was recently proved analytically by Motl
{\cite{motl}. The conjecture of Hod was within the framework
pioneered by Bekenstein in which the area spectrum is assumed to
be equally spaced \cite{beken}.
 The argument used by Hod and also by Dreyer goes as
follows: One assumes that  the relation between area and mass of a
non-rotation black hole is given by $A=16\pi M^2$. Its variations
are then $\Delta A= 32\pi M\,\Delta M$. If one assumes that the
variation in the mass is due to a quanta radiated with energy
$E_{\rm rad}=\hbar\, \omega_{\rm QNM}$, and uses the relation
(\ref{Hod}), one finds that the change in area is given by,
  \beq
      \Delta A=4\,\ln (3)\,l_{\rm P}^2\, .
  \eeq

Furthermore, Dreyer assumed that the change in area is due to an
appearance or disappearance of a puncture with spin $j_{\rm min}$.
Thus, one is lead to conclude that the Immirzi parameter is of the
form
 \beq
\gamma_{\rm d}=\frac{\ln(3)}{2\pi\sqrt{j_{\rm
      min}(j_{\rm min}+1)}}
\eeq
 Consistency with the Entropy calculation forces one
to take $j_{\rm min}=1$. Recall that the area contribution from an
edge with spin $j$ is given by $$A(j)=8\pi l^2_{\rm
P}\gamma\sqrt{j(j+1)},$$ and the entropy is given by
$$S=\frac{A}{4l^2_{\rm P}}\;\frac{\ln{(2j_{\rm
min}+1)}}{2\pi\gamma\,\sqrt{j_{\rm
      min}(j_{\rm min}+1)}}$$

As Dreyer recognizes, there are two possible attitudes one might
take:

(1) One assumes that the $j_{\rm min}=1$ is due to the minimum
possible value that $j$ can take, even at the kinematical level,
in which case one concludes that the gauge group should be
replaced by SO(3) (instead of SU(2));

(2) Think of something else.

Giving up the gauge group SU(2) is, at least to the author, an
undesirable step since one would loose the ability of the theory
to incorporate fermions. The purpose of this note is to propose a
physical argument within loop quantum gravity, that allows to keep
SU(2) as the gauge group, and at the same time have a consistent
description with the results of \cite{dreyer}. In fact, our
argument is strongly based on the requirement that fermions are
contained in LQG. In the following we shall assume that fermions
are included and the gauge group is SU(2).

This note is organized as follows. In Sec~\ref{sec:2}, we present
our argument within the loop quantum gravity formalism. In
Sec.~\ref{sec:3}, we reconsider the argument of quasinormal modes
in view of our conclusion of Sec.~\ref{sec:2}, and discuss the
implication for the loop quantum gravity program.

\section{Loop Quantum Gravity}
\label{sec:2}

In this section we shall focus our attention entirely to the LQG
formalism. Even when the following considerations are motivated by
Dreyer's results, from the logical point of view, they are
independent. Let us consider the physical process that would give
rise to the QNM frequency \cite{dreyer}: an appearance or
disappearance of a puncture with spin $j_{\rm min}$. We can think
of this process a being responsible, for instance, for the growth
of the BH when an edge that was ``free" gets attached to the
horizon. The inverse process could occur, say, when the horizon
gets excited and then ``emitts" \footnote{As it has already being
argued, this mechanism might not be the one responsible for
Hawking radiation, where the frequency spectrum is expected to be
different \cite{lee,motl}.}.

Now, in the process of disappearance of the puncture, this edge
becomes an open edge in the bulk. If the label of the edge were
$j=1/2$, then the only way to make the resulting state gauge
invariant is to have a fermion sitting at the end of the open
edge. However, this process would violate fermion conservation!
One could argue that at the same ``time", another similar process
takes place on the horizon such that fermion number is conserved.
However, a simpler attitude is to ask for {\it local} conservation
of fermion number. Thus, $j_{\rm min}=1/2$ is forbidden. The
minimum allowed value for the ``spin" of the resulting free edge
is $j_{\rm min}=1$. In that case, a pair of fermion-antifermion
could be attached to the end of the free edge respecting gauge
invariance and fermion number conservation. From this perspective,
the attachment and dis-attachment to the horizon of edges with
$j=1/2$ is a ``forbidden transition", and $j=1$ is the minimum
allowed value of $j$ that can puncture the black hole horizon, and
be responsible for the dynamical process of edge emission and
adsorption. Note also that, if this picture is correct, one can
only have integer values for $j$ touching the horizon.  Thus, even
when $j=1/2$ is allowed kinematically, it is the dynamical
consistency of the model that restricts the minimum $j$ involved
in the ``bulk-surface interaction".

Given that we are also considering the reverse process, namely
when a free edge attaches to the horizon, a process that could be
responsible for the black hole acquiring area (and mass). Then one
could heuristically expect that most of the area of the black hole
comes from this type of process, if the black hole was formed by a
dynamical process (as opposed to being an eternal black hole).

A possible conclusion of this argument is that in this new
picture, the main contribution to the entropy comes from $j_{\rm
min}=1$, since these would dominate over the $j=1/2$ edges. If
that is the case, then one is lead to conclude that the value of
the Immirzi parameter is the one consistent with this value,
namely $\gamma_{\rm d}=\frac{\ln(3)}{2\pi\sqrt{2}}$. Recall that
this parameter is a free parameter of the theory. If LQG is going
to be physically viable, then one is allowed to make a choice of
$\gamma$ only once. This choice has to be consistent with other
situations, or ``experiments". Since the computation of the black
hole entropy when matter is included (Maxwell, dilaton,
non-minimally coupled scalar field), as well as in more general
geometrical scenarios, such as for distorted and rotating horizons
\cite{abhay}, also yields {\it the same value of} $\gamma$, we
still have a consistent theory.

This is the main observation of this note.

{\it Remark}. Even when a detailed picture (i.e., a Hamiltonian)
for the dynamical situation here considered is still to be
constructed, the main principles behind our argument, namely gauge
invariance and fermion number conservation should be satisfied by
{\it any} such description. In this sense, our argument seems to
be robust.

\section{Discussion and Outlook}
\label{sec:3}

In the previous section we have argued that the existence of
fermions, together with a simple mechanism for adsorption and
emission of edges is consistent the the minimum value of $j=1$, we
can now return to the argument by Dreyer. In that case, the change
in the area that comes in the thermodynamic argument, complemented
with Bohr's correspondence principle, is consistent with the
(dis-)appearance of a $j=1$ edge.

Needless to say, this is a somewhat heuristic picture, and some
issues remain to be solved. For instance:
\begin{enumerate}

\item The existence of $j=1/2$ edges puncturing the horizon is not
forbidden (they would be something like ``primordial punctures"),
but they must be suppressed. Thus, one needs a dynamical
explanation of how exactly the entropy contribution is dominated
by the edges with the dynamical allowed value, namely $j=1$.

\item The derivation due to Hod and Dreyer is based on the
assumption that one has large black holes for which the entropy is
proportional to area and for which the thermodynamical relation
between area and mass is valid. One would like to have a complete
and systematic understanding of the picture at ``small" scales,
that is, near the Planck regime.

\item The heuristic physical process of conversion of area
quanta to matter quanta via the ``emission of an edge" is, of
course, very rough. One would like to have a clear picture of this
geometry-matter transition.

\item The existence of a universal limit for quasinormal modes
frequencies (depending only on the macroscopic parameters of the
BH) for non-rotating uncharged black holes is a remarkable fact.
An obvious question is whether such a dependence exists for
charged and rotating black holes, and if in that case, there is a
physical picture within LQG that can produce such frequencies.

\item Closely related to the previous point is the question of
whether one should be able to reproduce {\it all} possible QNM
frequencies (such as overtones) from allowed ``transitions" in LQG
(such as the emission of two edges). This in particular would
involve a departure from a ``Bohr correspondence principle" to a
detailed spectroscopy, which requires a precise understanding of
points (3) and (4) above.

\end{enumerate}

The Black Hole entropy calculation is amazing since it combines
many nontrivial facts about Chern Simons theory, loop quantum
geometry and thermodynamics. If we add the requirement that
fermions should be present, we have the highly unexpected result
that the (asymptotic) QNM frequencies, that know nothing about
$\hbar$ and quantum mechanics, are given by a simple physical
process of ``edge emission".

In this note we have presented an argument within LQG that
supports the minimum value of $j$ responsible for the BH entropy
to be $j_{\rm min}=1$, and thus making LQG consistent with the QNM
frequencies. Indeed, one could even argue for a stronger result.
Namely, one could say that a consistent framework for LQG,
incorporating fermions and black holes, requires $j_{\rm min}=1$
and in fact, {\it predicts} the QNM frequencies.

   The emerging physical picture is slightly changed
from the original considerations of ``it from bit" \cite{abck}
since the fundamental ``quanta" that give rise to entropy come
from spin 1 contributions, as opposed to spin 1/2. A full
understanding of these issues is then a matter of most importance.


\begin{acknowledgments}

I would like to thank A. Ashtekar and J.D. Bekenstein for
comments. This work was partially supported by a DGAPA-UNAM grant
No. IN112401 and a CONACyT grant No. J32754-E.

\end{acknowledgments}

\end{document}